\documentclass[letterpaper]{article} 
\usepackage{aaai2027}  
\usepackage[hyphens]{url}  
\usepackage{graphicx} 
\usepackage{natbib}  
\usepackage{caption} 
\usepackage{algorithm}
\usepackage{amsmath}
\usepackage{multirow} 
\usepackage{comment}
\usepackage{amssymb}
\usepackage{booktabs} 
\usepackage{array}
\usepackage{newfloat}
\usepackage{listings}
\DeclareCaptionStyle{ruled}{labelfont=normalfont,labelsep=colon,strut=off} 
\floatstyle{ruled}
\newfloat{listing}{tb}{lst}{}
\floatname{listing}{Listing}

\usepackage[export]{adjustbox}
\usepackage{rotating}
\usepackage{makecell}
\usepackage{tabularray}

\usepackage{booktabs}

\newif\ifshowrevisions
\showrevisionstrue   

\nocopyright 

\title{What Did the MLLM Hear? Token-Level Spectro-Temporal Grounding for Audio MLLM Explainability}
\author{
    Lucia Cascone, Valeria Fraenza\corresponding, Michele Nappi, Fabio Narducci, Benedetto Simone\\
}
\affiliations{

    \textit{Department of Computer Science, University of Salerno, Italy}\\
    lcascone@unisa.it, vfraenza@unisa.it, mnappi@unisa.it, fnarducci@unisa.it, bsimone@unisa.it
}

\begin{document}

\maketitle

\begin{abstract}
Audio-based Multimodal Large Language Models (MLLMs) can generate detailed natural-language descriptions of complex acoustic scenes, yet it remains unclear which parts of the input audio support each generated token. This is particularly challenging because acoustic evidence is distributed across time and frequency, and concurrent sound events may overlap temporally while occupying different spectral regions. We introduce STAG, to our knowledge the first post-hoc framework for token-level spectro-temporal grounding of captions generated by audio-based MLLMs. STAG estimates the temporal support for each generated token using target-token-specific vocabulary projections of the encoded audio representations, measures frequency-band relevance through controlled spectral occlusion, and combines the two signals into a spectro-temporal relevance map. We evaluate STAG against ten post-hoc explanation methods across four grounding benchmarks, where it achieves the best event-localization performance on every dataset, and apply it to eight audio-language backbones without parameter updates. Counterfactual deletion further shows that removing the identified evidence selectively reduces confidence in the corresponding event and frequently removes it from the regenerated caption. These results provide behavioral support for the faithfulness and selectivity of the explanations.
\end{abstract}

\section{Introduction}

Audio-based Multimodal Large Language Models (MLLMs) generate fluent,
open-vocabulary descriptions of complex acoustic scenes. However, a plausible caption
does not establish that each generated concept is supported by the input audio:
predictions may also reflect linguistic context, training correlations, or
language priors. Understanding these models therefore requires identifying
\emph{which acoustic evidence supports each generated token}. This problem is inherently token-specific. 
Content words may refer to different
events, sources, or acoustic attributes, whereas function words may require
little direct acoustic grounding. Moreover, relevant evidence is distributed
jointly over time and frequency. Temporal explanations identify when evidence
occurs, but cannot separate concurrent events occupying different spectral
regions; frequency-only explanations cannot determine when that evidence is
present. Most audio explainability methods target discriminative models and explain fixed
class scores \cite{dang2024explainableinterpretablemultimodallarge}. Audio MLLMs instead generate
autoregressive token sequences from latent audio representations. Existing
attribution and relevance-propagation methods may identify influential encoded
positions, but generally do not recover token-specific evidence over both time
and frequency.

\begin{figure*}[t]
    \centering
    \includegraphics[width=\textwidth]{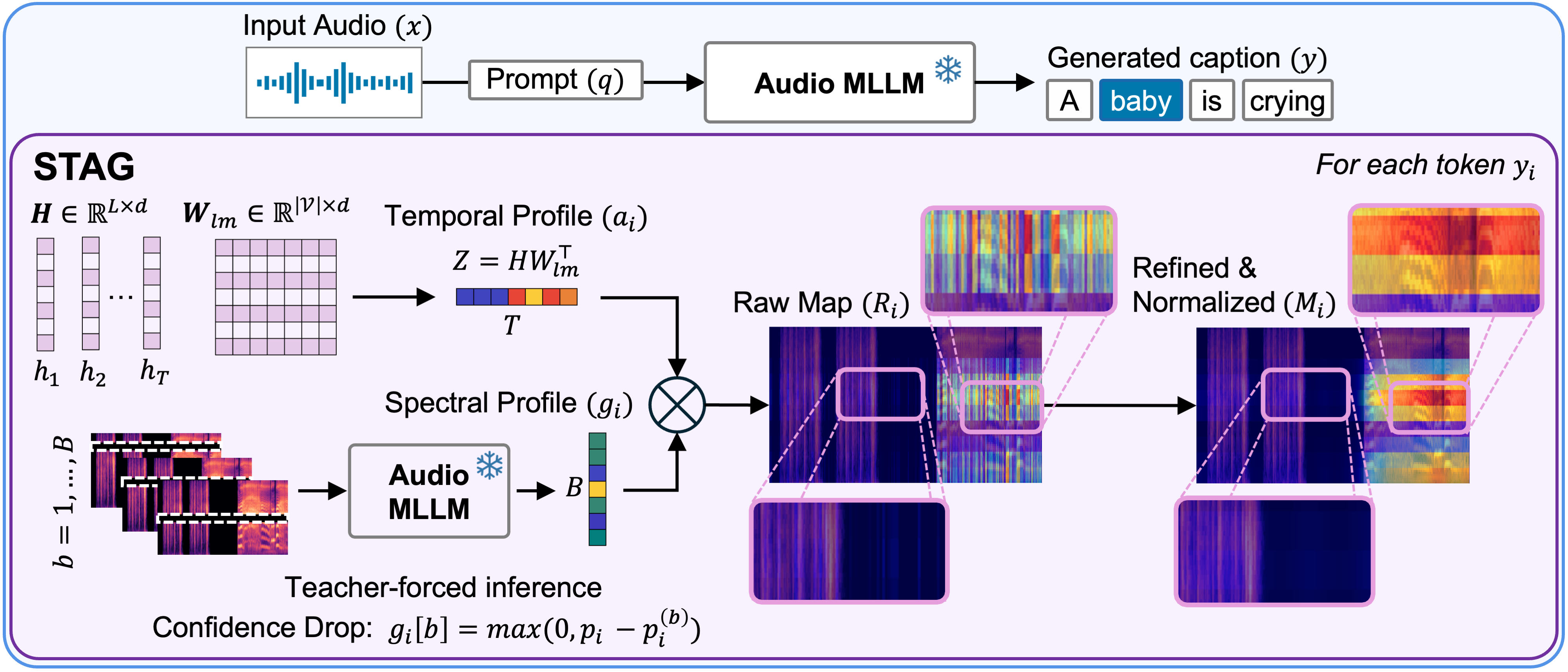}
    \caption{Overview of \textsc{STAG}. Token-conditioned temporal relevance and
    Frequency-Band Occlusion scores are fused into token-specific
    spectro-temporal maps, which can be aggregated into event-level
    explanations.}
    \label{fig:workflow}
\end{figure*}

We introduce \textbf{STAG}, a post-hoc framework for token-level
\textbf{S}pectro-\textbf{T}emporal \textbf{A}udio \textbf{G}rounding. For each
generated token, STAG derives temporal relevance from token-conditioned internal
activations and estimates spectral relevance through Frequency-Band Occlusion (FBO),
measuring the target-token probability decrease while keeping the preceding
sequence fixed. Their fusion yields a token-specific spectro-temporal map
without retraining or modifying the underlying model (Figure~\ref{fig:workflow}). Token maps referring to
the same acoustic event are subsequently aggregated for event-level analysis. We evaluate plausibility through temporal agreement with annotated sound events
and low relevance assigned to non-acoustic function words. We further assess
faithfulness using targeted counterfactual deletion, testing whether removing
the identified evidence selectively reduces the corresponding token probability
and suppresses the event in regenerated captions. Our main contributions are:
\begin{itemize}
    \item We introduce, to our knowledge, the first post-hoc framework for
    token-level spectro-temporal grounding of free-form captions generated by
    audio MLLMs, with token-to-event aggregation.
    \item We combine activation-based temporal attribution with
    Frequency-Band Occlusion, requiring only $B$ perturbed forward passes and
    no retraining or architectural modification.
    \item We compare \textsc{STAG} with ten post-hoc baselines on four grounding
    benchmarks, achieving the best event localization on every dataset, and
    demonstrate applicability to eight audio-language models. Counterfactual
    deletion further supports the faithfulness and selectivity of the
    explanations.
\end{itemize}

\section{Related Work}
Post-hoc explainability for audio has largely focused on discriminative models
and tasks other than open-vocabulary captioning. To the best of our knowledge,
no prior method provides token-level spectro-temporal explanations for audio
MLLMs. Our comparison therefore considers general-purpose attribution methods
adapted to audio-language generation and token-conditioned localization methods
originally developed for vision-language models. General-purpose attribution methods can be adapted to audio-language models by
assigning relevance to encoded audio positions. Gradient-based localization
includes Grad-CAM \cite{selvaraju2017grad}, Grad-CAM++ \cite{8354201}, and
LayerCAM \cite{jiang2021layercam}. Attention-based methods use raw attention or
aggregate it across layers, optionally incorporating output gradients
\cite{abnar-zuidema-2020-quantifying}. AttnLRP and CP-LRP instead propagate
relevance through Transformer components
\cite{achtibat2024attnlrp,ali2022xai}, while Architectural Surgery intervenes on
internal multimodal representations to improve localization
\cite{LI2025111409}. These methods constitute the main baseline families in our
evaluation. When applied to encoded audio sequences, however, they provide
relevance along the temporal representation without explicitly resolving its
frequency support. Token-conditioned localization has also been studied in vision-language models.
TAM \cite{Li_2025_ICCV} projects visual features through token-specific output
weights and refines the resulting maps using Estimated Causal Inference (ECI) and
Rank Gaussian (RG) filtering. While TAM localizes generated tokens over a spatial
image grid, audio-caption grounding requires resolving evidence jointly across
time and frequency. \textsc{STAG} addresses this setting by combining
token-conditioned temporal attribution with FBO.


\section{Method}
\label{sec:method}

Given a waveform $x$, a prompt $q$, and a generated caption $\mathbf{y}=(y_1,\ldots,y_N)$, \textsc{STAG} produces for each token $y_i$ a non-negative relevance map $\mathbf{M}_i\in\mathbb{R}_{\geq 0}^{B\times T}$ over $B$ frequency bands and $T$ encoded audio positions. As illustrated in Figure~\ref{fig:workflow}, the map combines a temporal profile derived from internal audio representations with a frequency profile estimated through controlled spectral occlusion. Their factorized fusion, followed by temporal refinement and
shared normalization, yields a token-specific spectro-temporal explanation
without retraining the underlying model.

\subsection{Token-Specific Temporal Attribution}
\label{sec:temporal-attribution}
 Let $\mathbf{H}\in\mathbb{R}^{L\times d}$ denote the final-layer hidden states obtained during multimodal prefill, where $T$ of the $L$ positions correspond to encoded audio features. Applying the frozen language-modeling head
$\mathbf{W}_{\mathrm{lm}}\in\mathbb{R}^{|\mathcal{V}|\times d}$ yields
position-wise vocabulary scores $\mathbf{Z}=\mathbf{H}\mathbf{W}_{\mathrm{lm}}^{\top}$.
With $c_i$ the vocabulary index of $y_i$, selecting the target-token score at
each audio position defines the temporal profile:
\begin{equation}
a_i[t]=\left[Z_{\pi(t),c_i}\right]_{+},\qquad t=1,\ldots,T,
\label{eq:temporal-profile}
\end{equation}
where $[z]_{+}=\max(0,z)$ and $\pi(t)$ maps the $t$-th audio position to its
absolute index in the sequence. As encoded audio positions preserve temporal
order, $\mathbf{a}_i\in\mathbb{R}_{\geq 0}^{T}$ is an ordered profile of positive associations with $y_i$.

\subsection{Frequency-Band Occlusion}
\label{sec:fbo}

In polyphonic audio, overlapping events may occupy different frequency
regions. \textsc{STAG} therefore introduces \emph{Frequency-Band Occlusion}
(FBO), a deterministic black-box procedure that partitions
$[0,\mathrm{sr}/2]$ into $B$ contiguous Mel-spaced bands
$[f_{b-1},f_b)$, where 
$\mathrm{sr}/2$ is the Nyquist frequency.

Let $\mathbf{X}$ be the STFT of waveform $x$, with $X[k,\tau]$ denoting the
coefficient at frequency bin $k$ and frame $\tau$, and $\nu_k$ its physical
frequency. Band $b$ is removed across all frames as $\widetilde{X}^{(b)}[k,\tau]=X[k,\tau]\,1\!\left[\nu_k\notin[f_{b-1},f_b)\right],$
where $1[\cdot]$ is the indicator function. The perturbed waveform
$x^{(b)}$ is reconstructed by inverse STFT. Since temporal localization is
provided by $\mathbf{a}_i$, FBO occludes each band over the full recording and
requires only $B$ perturbed forward passes instead of $BT$ time-frequency
perturbations. Each $x^{(b)}$ is evaluated under teacher forcing with prompt $q$ and prefix
$\mathbf{y}_{<i}=(y_1,\ldots,y_{i-1})$ fixed. Let $p_{\theta}$ denote the
next-token probability of the frozen model, with
$p_i=p_{\theta}(y_i\mid x,q,\mathbf{y}_{<i})$ and
$p_i^{(b)}=p_{\theta}(y_i\mid x^{(b)},q,\mathbf{y}_{<i})$. The spectral profile
$\mathbf{g}_i\in\mathbb{R}_{\geq 0}^{B}$ is:
\begin{equation}
g_i[b]
=
\max\!\left(0,p_i-p_i^{(b)}\right),
\qquad b=1,\ldots,B.
\label{eq:spectral-profile}
\end{equation}
Thus, a band is relevant only when its removal decreases the probability of
$y_i$. The same $B$ teacher-forced passes provide spectral relevance for all
generated tokens.

\subsection{Spectro-Temporal Fusion and Refinement}
\label{sec:fusion}

\paragraph{Factorized fusion.}
The rectified temporal and spectral profiles are fused directly, without prior
normalization:
\begin{equation}
\mathbf{R}_i
=
\mathbf{g}_i\otimes\mathbf{a}_i,
\qquad
R_i[b,t]
=
g_i[b]\,a_i[t],
\label{eq:spectro-temporal-fusion}
\end{equation}
where $\mathbf{R}_i\in\mathbb{R}_{\geq 0}^{B\times T}$. The resulting map has
rank at most one and assumes separability across time and frequency. Although it
cannot represent arbitrary local time-frequency interactions, polyphonic scenes
can be described through multiple token-specific maps: overlapping events may
share similar temporal profiles while retaining different spectral profiles.

\paragraph{Temporal refinement.}
To reduce fragmentation across adjacent audio positions, each frequency band is
smoothed independently along time:
\begin{equation}
\widetilde{R}_i[b,t]
=
\sum_{s=1}^{T}
R_i[b,s]\,
G_{\sigma_{\mathrm{time}}}(t-s),
\label{eq:temporal-refinement}
\end{equation}
where $G_{\sigma_{\mathrm{time}}}$ is a normalized one-dimensional Gaussian
kernel with temporal bandwidth $\sigma_{\mathrm{time}}$. Smoothing is not
applied across frequency, since the bands are estimated through independent
occlusions.

\paragraph{Shared normalization.}
The refined maps of all tokens from the same sample are normalized by a shared
maximum:
\begin{equation}
M_i[b,t]
=
\frac{\widetilde{R}_i[b,t]}
{\displaystyle
\max_{j,b',t'}
\widetilde{R}_j[b',t']
+\epsilon},
\label{eq:global-normalization}
\end{equation}
where $j$ indexes generated tokens and $\epsilon>0$ ensures numerical
stability. Shared normalization preserves relative relevance across tokens, so
weakly attributed tokens remain weak. The resulting
$\mathbf{M}_i\in\mathbb{R}_{\geq 0}^{B\times T}$ is the final token-specific
spectro-temporal explanation.

\subsection{Token-to-Event Aggregation}
\label{sec:aggregation}

Subtokens are merged into words, and words referring to the same acoustic event
are grouped using the dependency-based procedure described in the supplementary
material. Let $\mathcal{I}_E$ denote the token indices assigned to event $E$.
Their maps are aggregated elementwise:
$\mathbf{M}_E[b,t]
=
\max_{i\in\mathcal{I}_E} M_i[b,t].
\label{eq:event-map-aggregation}
$
Unlike summation, maximum aggregation does not inflate relevance when an event
spans multiple tokens and preserves
$\mathbf{M}_E\in[0,1]^{B\times T}$. \textsc{STAG} therefore provides
token-level maps $\mathbf{M}_i$ and event-level maps $\mathbf{M}_E$, both
retaining the full $B\times T$ spectro-temporal structure.

\begin{figure*}[t!]
    \centering
    \includegraphics[width=1\linewidth]{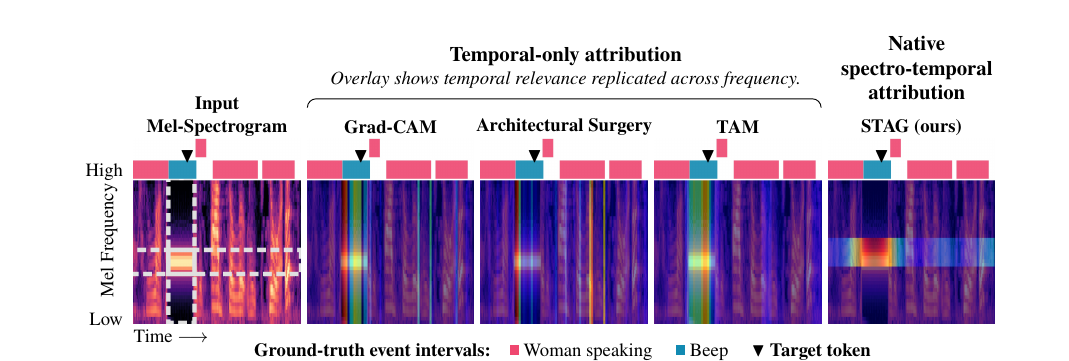}
    \caption{Attribution maps for the target token \emph{beep}. The temporal
profiles produced by comparison methods are replicated
across frequency for visualization, whereas \textsc{STAG} directly produces a
spectro-temporal relevance map. Dashed guides mark the visually identifiable
time--frequency region of the beep and are not benchmark annotations.}
    \label{fig:ComparisonSOA_Short}
\end{figure*} 

\section{Experimental Evaluation}
\label{sec:experiments}

\subsection{Experimental Setup}

\subsubsection{Datasets.}
We evaluate \textsc{STAG} on the official test splits of four complementary
benchmarks with
temporal onset-offset annotations: AudioTime \cite{xie2025audiotime}
(500 controlled clips, two ordered events each), AudioGrounding
\cite{xu2021text} (492 real-world polyphonic recordings with phrase-level
intervals), TACoS \cite{primus2025tacos} (2{,}000 longer recordings with dense
time-aligned captions), and AudioSet-Strong \cite{hershey2021benefit}
(14{,}045 polyphonic clips with class-level boundaries).

\subsubsection{Audio-Language Models.}
We evaluate \textsc{STAG} on eight publicly available models spanning diverse
audio encoders, fusion strategies, and backbones: Qwen2.5-Omni-3B/7B
\cite{xu2025qwen25omnitechnicalreport}, Qwen2-Audio-7B and -Instruct
\cite{qwen2audio}, Kimi-Audio-7B-Instruct \cite{kimiaudio}, Audio Flamingo 3
\cite{audioflamingo}, Audio Flamingo Next
\cite{ghosh2026audioflamingonextnextgeneration}, and Gemma 4-12B-Instruct
\cite{gemmateam2026gemma4}. All are evaluated without parameter updates using
the prompt \textit{``Write a one-sentence caption for this sound:''}. Because
each model generates its own caption, the extracted events and their
ground-truth matches vary across models. Each model is integrated through a
lightweight runner exposing the token-level scores and next-token probabilities
required by \textsc{STAG}; all subsequent attribution, fusion, aggregation, and
evaluation stages are shared. Model-specific adaptations are given in the
supplementary material.

\subsubsection{Implementation Details.}
Audio is resampled to 16 kHz. FBO uses an STFT with a 400-sample window,
160-sample hop, and $B=10$ Mel-spaced bands.  Temporal refinement uses
$\sigma_{\mathrm{time}}=6$ encoded audio positions. Event matching uses WordNet
\cite{miller-1994-wordnet}, followed, when necessary, by
\texttt{all-MiniLM-L6-v2} sentence embeddings
\cite{wang2020minilm} with a cosine-similarity threshold of $0.8$.
Additional implementation details and hyperparameter analyses are provided in
the supplementary material.

\subsection{Evaluation Protocol}

Throughout the qualitative figures, colored bars above the spectrograms denote
the benchmark ground-truth temporal intervals, with colors corresponding to the
event labels shown in each figure. Text below each spectrogram reports the
corresponding model output or the relevant excerpt from its generated caption.

\subsubsection{Motivating Example.}
\label{sec:motivating-example}

Figure~\ref{fig:ComparisonSOA_Short} illustrates the token-specific
spectro-temporal relevance produced by \textsc{STAG}. The first panel shows the
input Mel-spectrogram. Dashed vertical and horizontal guides mark, respectively,
the temporal extent and visually identifiable frequency band of the synthetic
beep associated with the target token \emph{beep}; they are included only for
reference and are not benchmark annotations. This example is selected because
the beep has a compact and clearly distinguishable spectral signature. The remaining panels compare explanations for the same target token. The
baseline methods provide temporal relevance only, whereas \textsc{STAG}
produces a two-dimensional map over time and frequency. Since the benchmarks
provide temporal boundaries but no frequency-level annotations, spectral
localization is assessed qualitatively, while quantitative evaluation requires
projecting the relevance maps onto time.

\subsubsection{Temporal Projection.}
\label{sec:projection}

To compare with the available temporal annotations, each refined token map
$\widetilde{\mathbf{R}}_i\in\mathbb{R}_{\geq 0}^{B\times T}$ is projected onto
time.  We define
$e_i[b]=\sum_{t=1}^{T}\widetilde{R}_i[b,t]^2$ and
$w_i[b]=e_i[b]/\sum_{b'=1}^{B}e_i[b']$, using $w_i[b]=1/B$ when the map is
identically zero. The projected profile is
\begin{equation}
\overline{m}_i[t]
=
\sum_{b=1}^{B}
w_i[b]\widetilde{R}_i[b,t].
\label{eq:raw-temporal-projection}
\end{equation}
Profiles from the same sample are jointly normalized by their shared maximum,
yielding $\mathbf{m}_i\in\mathbb{R}_{\geq 0}^{T}$. The event-level profile is
$m_E[t]=\max_{i\in\mathcal{I}_E}m_i[t]$. The profile $\mathbf{m}_E$ is used only for temporal grounding metrics and is
distinct from the spectro-temporal map $\mathbf{M}_E$: $\mathbf{m}_E$ projects token maps before
aggregation, whereas $\mathbf{M}_E$ aggregates them directly in the
$B\times T$ domain.

\subsubsection{Grounding Metrics.} \textbf{Event localization (E).}
Generated events are aligned with ground-truth labels through lemma matching,
followed by synonym and sentence-embedding matching when needed. The large-scale
XAI comparison uses lexical matching only to ensure an identical protocol across
methods. For each matched event $E\in\mathcal{E}_{\mathrm{match}}$, its
annotated intervals are merged into a binary mask
$\mathbf{G}_E\in\{0,1\}^{T}$. The temporal profile $\mathbf{m}_E$ is
independently re-normalized by min-max scaling to restore its per-event dynamic
range for Otsu thresholding, yielding the binary prediction
$\mathbf{P}_E$. Event localization is
\begin{equation}
\mathrm{Event\text{-}tIoU}(E)
=
\frac{|\mathbf{P}_E\cap\mathbf{G}_E|}
     {|\mathbf{P}_E\cup\mathbf{G}_E|}.
\label{eq:event-tiou}
\end{equation}
The reported score is averaged over $\mathcal{E}_{\mathrm{match}}$ and is
therefore conditional on generating an alignable event.

\paragraph{Function-word silence (S).}
We define a sample-specific threshold $\bar{\theta}$ by averaging the Otsu
thresholds of valid event profiles on the shared per-sample normalization scale,
where event and function-word profiles remain directly comparable. For each
function word, we compute the fraction of temporal positions below
$\bar{\theta}$. \emph{Func-Silence} (S) is averaged over evaluable function
words; samples without valid events or function words are excluded.

\paragraph{Combined metric (F1).}
\emph{Event-F1} (F1) is the harmonic mean of E and S, penalizing methods that
perform well on only one criterion.

\subsubsection{Counterfactual Faithfulness Analysis.}
\label{sec:counterfactual}

Temporal agreement does not establish that the model relies on the identified
evidence. For each valid event $E$, we binarize its map $\mathbf{M}_E$ using
Otsu's method and resize the mask to the STFT grid,
$\mathbf{\Omega}_E\in\{0,1\}^{K\times\mathcal{T}}$, where $K$ and
$\mathcal{T}$ denote STFT frequency bins and temporal frames. We remove the
selected coefficients as
$X_E^{\mathrm{del}}[k,\tau]
=
X[k,\tau](1-\Omega_E[k,\tau])$
and reconstruct the counterfactual waveform $x_E^{\mathrm{del}}$ by inverse
STFT.

\paragraph{Target confidence.}
Under teacher forcing, we measure the mean log-probability change over the
tokens assigned to $E$:
\begin{equation}
\begin{split}
\Delta_{\mathrm{tgt}}(E)
=
\frac{1}{|\mathcal{I}_E|}
\sum_{i\in\mathcal{I}_E}
\Big[
&\log p_{\theta}
\!\left(y_i\mid x_E^{\mathrm{del}},q,\mathbf{y}_{<i}\right)
\\
-&\log p_{\theta}
\!\left(y_i\mid x,q,\mathbf{y}_{<i}\right)
\Big].
\end{split}
\label{eq:target-confidence-drop}
\end{equation}
Negative values indicate reduced target-event confidence. We report the mean
$\Delta\log p$ with its 95\% bootstrap confidence interval and
\emph{Conf.\ Drop}, the fraction of events satisfying
$\Delta_{\mathrm{tgt}}(E)<-\ln 2$.

\paragraph{Specificity and disappearance.}
Under the same counterfactual input, we compute the analogous change for every
non-target event $E'\neq E$. We denote by $\Delta_{\mathrm{non}}$ the mean
non-target change over all target--non-target pairs and report it as
\emph{Specificity}. Selectivity is supported when
$|\Delta_{\mathrm{non}}|<|\Delta_{\mathrm{tgt}}|$, indicating that the target
event is affected more strongly, although overlapping sources may produce
collateral effects. We also regenerate the caption without teacher forcing and
report \emph{Disappear}, the fraction of target events for which none of the
associated lemmas reappears in the regenerated caption. Mask area and removed
energy measure the fractions of selected STFT bins and removed STFT energy,
respectively.

\section{Experimental Results}
\label{sec:results}

\begin{table*}[t]
\centering
\setlength{\tabcolsep}{1mm}

{\small
\begin{tabular}{@{}lrrrrrrrrrrrrrrrr@{}}
\toprule
&
\multicolumn{4}{c}{\textbf{AudioTime}} &
\multicolumn{4}{c}{\textbf{AudioGrounding}} &
\multicolumn{4}{c}{\textbf{TACoS}} &
\multicolumn{4}{c}{\textbf{AudioSet-Strong}} \\
\cmidrule(lr){2-5}
\cmidrule(lr){6-9}
\cmidrule(lr){10-13}
\cmidrule(l){14-17}

\textbf{Model}
& $N_m$ & \textit{E}$\uparrow$ & \textit{S}$\uparrow$ & \textit{F1}$\uparrow$
& $N_m$ & \textit{E}$\uparrow$ & \textit{S}$\uparrow$ & \textit{F1}$\uparrow$
& $N_m$ & \textit{E}$\uparrow$ & \textit{S}$\uparrow$ & \textit{F1}$\uparrow$
& $N_m$ & \textit{E}$\uparrow$ & \textit{S}$\uparrow$ & \textit{F1}$\uparrow$ \\
\midrule

Omni-7B
& 596 & 50.63 & 79.31 & 61.81
& 947 & 41.17 & 78.32 & 53.97
& 3421 & 37.29 & 81.10 & 51.09
& 19856 & 37.04 & 76.84 & 49.99 \\

Omni-3B
& 502 & \textbf{52.09} & \textbf{89.99} & \textbf{65.99}
& 787 & \textbf{41.97} & \textbf{89.06} & \textbf{57.05}
& 3364 & 34.23 & \textbf{87.40} & 49.19
& 17826 & 37.64 & \textbf{84.60} & \textbf{52.10} \\

\midrule

Q2A-7B-I
& 422 & 21.13 & 85.98 & 33.93
& 725 & 31.17 & 82.14 & 45.19
& 2718 & 37.10 & 76.33 & 49.93
& 13737 & 30.39 & 81.05 & 44.20 \\

Q2A-7B
& 333 & 17.03 & 84.73 & 28.36
& 513 & 26.30 & 74.98 & 38.94
& 1945 & 34.35 & 72.55 & 46.63
& 10483 & 28.46 & 75.13 & 41.28 \\

\midrule

Kimi-7B-I
& 265 & 43.11 & 71.93 & 53.91
& 729 & 38.90 & 72.19 & 50.56
& 2471 & 39.50 & 74.57 & 51.64
& 13763 & \textbf{39.08} & 71.99 & 50.66 \\

\midrule

AF3
& 720 & 38.47 & 86.20 & 53.20
& 1224 & 36.98 & 85.31 & 51.60
& 4426 & \textbf{40.49} & 80.62 & \textbf{53.90}
& 36431 & 32.87 & 82.48 & 47.01 \\

AF-Next
& 636 & 14.45 & 85.31 & 24.72
& 1124 & 30.28 & 84.08 & 44.53
& 4779 & 37.99 & 84.11 & 52.34
& 29026 & 32.28 & 80.89 & 46.15 \\

\midrule

Gemma-12B-I
& 47 & 23.17 & 39.94 & 29.33
& 104 & 23.36 & 38.21 & 28.99
& 1270 & 18.68 & 47.89 & 26.88
& 4113 & 18.43 & 42.36 & 25.68 \\

\bottomrule
\end{tabular}
}

\caption{Cross-model evaluation of \textsc{STAG} using semantic event matching.
$N_m$ is the number of generated events matched to the annotations. Omni, Q2A, AF, and I
denote Qwen2.5-Omni, Qwen2-Audio, Audio Flamingo, and Instruct, respectively.}
\label{tab:models_comparison}
\end{table*}

\subsection{Cross-Model Applicability and Backbone Selection} Table~\ref{tab:models_comparison} evaluates \textsc{STAG} across eight
audio-language models. Since each model generates its own
caption, E, S, and F1 are computed over the generated events matched to the
ground-truth annotations. The results therefore characterize grounding behavior
rather than captioning quality and should be interpreted together with the
matched-event count $N_m$.

Qwen2.5-Omni-3B achieves the highest F1 on AudioTime, AudioGrounding, and
AudioSet-Strong, the highest S on all four benchmarks, and the highest E on
AudioTime and AudioGrounding. Audio Flamingo~3 ranks first on TACoS, whereas
Kimi-Audio-7B-Instruct attains the highest E on AudioSet-Strong but substantially
lower S. Grounding behavior is therefore both model- and benchmark-dependent.

\begin{figure}[t!]
    \centering
    \includegraphics[width=1\linewidth]{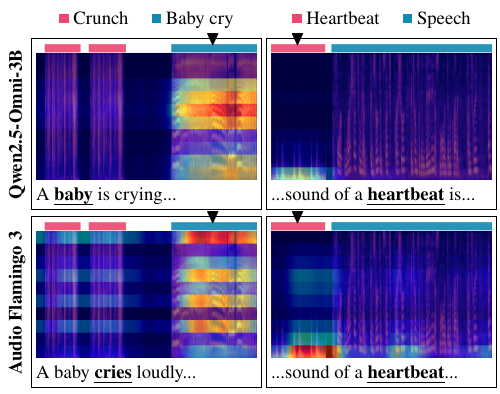}
    \caption{Representative cross-model comparison of \textsc{STAG} relevance
    maps for the same audio inputs. Although the models mention the same main events, their maps differ in the
spectral evidence and source emphasis.}
    \label{fig:stag_models}
\end{figure}

Qwen2.5-Omni-3B also outperforms its 7B counterpart on three of the four
benchmarks despite having fewer parameters. Given this favorable trade-off
between grounding performance and model size, we adopt it as the reference
backbone for the comparison with post-hoc baselines and for the subsequent analyses. Figure~\ref{fig:stag_models} complements these aggregate results by comparing
the relevance maps produced for the same audio input. In the first example, the
models mention the ``baby cry'' but assign relevance to different frequency regions
of the cry and the preceding ``crunch'' sound. In the second, all mention the
``heartbeat'' but differ in the relevance assigned to the subsequent speech,
yielding captions with different semantic emphasis. Thus, similar captions and
temporal agreement need not correspond to equivalent spectro-temporal
explanations. Additional examples are provided in the supplementary material.

\subsection{Comparison with Post-hoc Explainability Methods}
\label{sec:baselines}

We compare \textsc{STAG} with ten post-hoc explainability baselines using
identical Qwen2.5-Omni-3B captions and lexical event matching
(Table~\ref{tab:sota_comparison}). \textsc{STAG} achieves the highest E and F1
on all four benchmarks, outperforming the strongest baseline on each dataset
(TAM on AudioTime and Grad-CAM on the remaining benchmarks) by 7.58--12.84 E
points and 8.70--14.33 F1 points. Attention- and relevance-propagation methods
attain near-ceiling S but substantially lower E, indicating that suppressing
relevance for function words does not necessarily yield accurate event
localization. F1 captures this imbalance, whereas \textsc{STAG} combines strong
event localization with consistently high function-word silence. LayerCAM
produces flat relevance profiles that yield empty masks after Otsu thresholding
and consequently scores zero on all benchmarks.

\begin{table*}[ht]
\centering
\setlength{\tabcolsep}{1mm}

{\small
\begin{tabular}{@{}ll*{4}{rrr}@{}}
\toprule
&
& \multicolumn{3}{c}{\textbf{AudioTime}}
& \multicolumn{3}{c}{\textbf{AudioGrounding}}
& \multicolumn{3}{c}{\textbf{TACoS}}
& \multicolumn{3}{c}{\textbf{AudioSet-Strong}} \\
\cmidrule(lr){3-5}
\cmidrule(lr){6-8}
\cmidrule(lr){9-11}
\cmidrule(l){12-14}

\textbf{Family}
& \textbf{Method}
& \textit{E}$\uparrow$ & \textit{S}$\uparrow$ & \textit{F1}$\uparrow$
& \textit{E}$\uparrow$ & \textit{S}$\uparrow$ & \textit{F1}$\uparrow$
& \textit{E}$\uparrow$ & \textit{S}$\uparrow$ & \textit{F1}$\uparrow$
& \textit{E}$\uparrow$ & \textit{S}$\uparrow$ & \textit{F1}$\uparrow$ \\
\midrule

\multirow[c]{3}{*}{Gradient}
& Grad-CAM
& 33.68 & 79.56 & 47.33
& \underline{29.61} & 79.60 & \underline{43.16}
& \underline{30.02} & 80.66 & \underline{43.75}
& \underline{31.08} & 77.52 & \underline{44.37} \\

& Grad-CAM++
& 20.36 & 79.29 & 32.40
& 26.07 & 79.68 & 39.28
& 29.54 & 78.10 & 42.87
& 27.04 & 79.67 & 40.38 \\

& LayerCAM
& 0.00 & 0.00 & 0.00
& 0.00 & 0.00 & 0.00
& 0.00 & 0.00 & 0.00
& 0.00 & 0.00 & 0.00 \\

\midrule

\multirow[c]{3}{*}{Attention}
& Raw Attention
& 1.88 & \textbf{99.20} & 3.69
& 2.14 & \textbf{98.45} & 4.19
& 1.01 & \textbf{99.30} & 2.00
& 2.26 & \textbf{98.32} & 4.41 \\

& Attention Rollout
& 3.32 & 93.07 & 6.41
& 6.17 & 93.76 & 11.58
& 3.41 & 97.36 & 6.59
& 6.78 & 93.31 & 12.64 \\

& Gradient Rollout
& 17.19 & 96.65 & 29.18
& 9.55 & 96.84 & 17.38
& 5.99 & 97.93 & 11.29
& 6.57 & 96.74 & 12.31 \\

\midrule

\multirow[c]{2}{*}{Relevance}
& AttnLRP
& 5.52 & \underline{98.36} & 10.46
& 4.40 & \underline{97.87} & 8.42
& 2.61 & \underline{98.95} & 5.09
& 4.41 & \underline{97.82} & 8.43 \\

& CP-LRP
& 6.41 & 97.95 & 12.04
& 4.93 & 97.60 & 9.38
& 2.59 & 98.77 & 5.04
& 4.89 & 97.47 & 9.31 \\

\midrule

Architecture
& Architectural Surgery
& 22.23 & 92.04 & 35.81
& 13.78 & 92.91 & 24.00
& 9.10 & 95.36 & 16.62
& 11.39 & 92.35 & 20.27 \\

\midrule

Activation
& TAM
& \underline{42.00} & 85.19 & \underline{56.26}
& 27.00 & 82.63 & 40.70
& 24.65 & 82.03 & 37.90
& 23.09 & 80.81 & 35.92 \\

\midrule

\textbf{Ours}
& \textbf{\textsc{STAG}}
& \textbf{54.42} & 89.99 & \textbf{67.82}
& \textbf{42.45} & 89.06 & \textbf{57.49}
& \textbf{40.37} & 87.40 & \textbf{55.22}
& \textbf{38.66} & 84.60 & \textbf{53.07} \\

\bottomrule
\end{tabular}
}

\caption{Comparison with ten post-hoc explanation methods on four temporal
grounding benchmarks using identical Qwen2.5-Omni-3B captions and lexical event
matching. }
\label{tab:sota_comparison}
\end{table*}

\subsection{Component and Projection Analysis}
\label{sec:ablation}

Table~\ref{tab:ablation} compares the main component and projection
configurations of \textsc{STAG}, all evaluated using semantic event matching.
The temporal reference corresponds to the temporal-only attribution setting
used by TAM. Because it has no frequency axis and uses direct projection, its
comparison with the spectro-temporal configurations reflects the joint effect
of introducing frequency-band evidence and changing the projection strategy,
rather than an isolated contribution of FBO. Nevertheless, all
spectro-temporal configurations achieve higher F1 than the temporal reference,
supporting the complementarity of spectral and temporal evidence.

The remaining rows provide three controlled comparisons. With RG processing and
max projection fixed, removing ECI improves F1 on all four benchmarks,
indicating that inter-token subtraction is not beneficial in this setting. With
ECI disabled and max projection unchanged, replacing RG with TR yields the
largest improvement, increasing F1 by 7.26--8.48 points. Finally, max and
weighted projection differ by only 0.04--0.26 F1 points, indicating limited
sensitivity to the projection rule. We therefore adopt weighted projection in
the final \textsc{STAG} configuration.
Because all rows use semantic event matching, their values are not directly
comparable with the lexical-matching results in
Table~\ref{tab:sota_comparison}.

\begin{table*}[t]
    \centering
    \setlength{\tabcolsep}{1mm}

    {\small
    \begin{tabular}{@{}lcccc|ccc|ccc|ccc|ccc@{}}
        \toprule
        \multirow{2}{*}{\textbf{Configuration}}
        & \multirow{2}{*}{\textbf{FBO}}
        & \multirow{2}{*}{\textbf{ECI}}
        & \multirow{2}{*}{\textbf{Ref.}}
        & \multirow{2}{*}{\textbf{Proj.}}
        & \multicolumn{3}{c|}{\textbf{AudioTime}}
        & \multicolumn{3}{c|}{\textbf{AudioGrounding}}
        & \multicolumn{3}{c|}{\textbf{TACoS}}
        & \multicolumn{3}{c}{\textbf{AudioSet-Strong}} \\
        \cmidrule(lr){6-8}
        \cmidrule(lr){9-11}
        \cmidrule(lr){12-14}
        \cmidrule(l){15-17}

        & & & &
        & \textit{E}$\uparrow$ & \textit{S}$\uparrow$ & \textit{F1}$\uparrow$
        & \textit{E}$\uparrow$ & \textit{S}$\uparrow$ & \textit{F1}$\uparrow$
        & \textit{E}$\uparrow$ & \textit{S}$\uparrow$ & \textit{F1}$\uparrow$
        & \textit{E}$\uparrow$ & \textit{S}$\uparrow$ & \textit{F1}$\uparrow$ \\
        \midrule

        Temporal ref.
        & $\times$ & $\checkmark$ & RG & D
        & 40.60 & 85.19 & 54.99
        & 26.50 & 82.63 & 40.13
        & 21.37 & 82.03 & 33.91
        & 22.88 & 80.81 & 35.67 \\

        FBO + ECI
        & $\checkmark$ & $\checkmark$ & RG & M
        & 41.40 & 89.83 & 56.68
        & 30.77 & \textbf{89.59} & 45.80
        & 24.72 & 87.07 & 38.50
        & 27.33 & \textbf{87.12} & 41.61 \\

        FBO, no ECI
        & $\checkmark$ & $\times$ & RG & M
        & 42.52 & 89.81 & 57.72
        & 33.31 & 88.69 & 48.43
        & 27.50 & 86.98 & 41.79
        & 29.79 & 85.96 & 44.24 \\

        FBO + TR
        & $\checkmark$ & $\times$ & TR & M
        & 51.93 & 89.53 & 65.73
        & 41.89 & 88.76 & 56.91
        & 34.14 & 87.09 & 49.05
        & 37.61 & 84.58 & 52.06 \\

        \midrule

        \textbf{\textsc{STAG}}
        & $\checkmark$ & $\times$ & TR & W
        & \textbf{52.09} & \textbf{89.99} & \textbf{65.99}
        & \textbf{41.97} & 89.06 & \textbf{57.05}
        & \textbf{34.23} & \textbf{87.40} & \textbf{49.19}
        & \textbf{37.64} & 84.60 & \textbf{52.10} \\

        \bottomrule
    \end{tabular}
    }
\caption{Comparison of \textsc{STAG} component and projection configurations
using Qwen2.5-Omni-3B. D, M, and W denote direct, max, and weighted temporal projection,
respectively.}
    \label{tab:ablation}
\end{table*}

\subsection{Label Ambiguity and Matching Limitations}
\label{sec:confounding}

\begin{figure}[t]
    \centering
    \includegraphics[width=\linewidth]
    {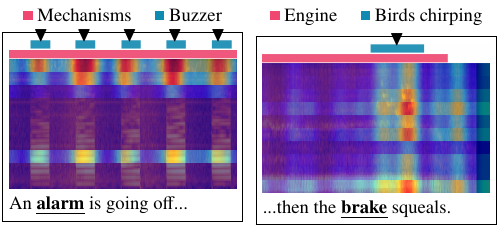}
    \caption{Examples of label ambiguity for Qwen2.5-Omni-3B. The generated
    event labels differ from the benchmark annotations but remain acoustically
    plausible, while their \textsc{STAG} maps overlap the annotated temporal
    intervals.}
    \label{fig:confounding_sounds}
\end{figure}

Figure~\ref{fig:confounding_sounds} illustrates cases in which acoustically
similar events receive different but perceptually plausible labels. Although
the generated event is not matched to the benchmark annotation, its
\textsc{STAG} map overlaps the annotated temporal interval. These examples show
that some apparent matching failures may arise from label ambiguity rather than
incorrect acoustic localization. Token-level explanations can therefore reveal
grounding behavior that is not captured by aggregate matching-based metrics.

\begin{figure*}[t!]
    \centering
    \includegraphics[width=1\linewidth]{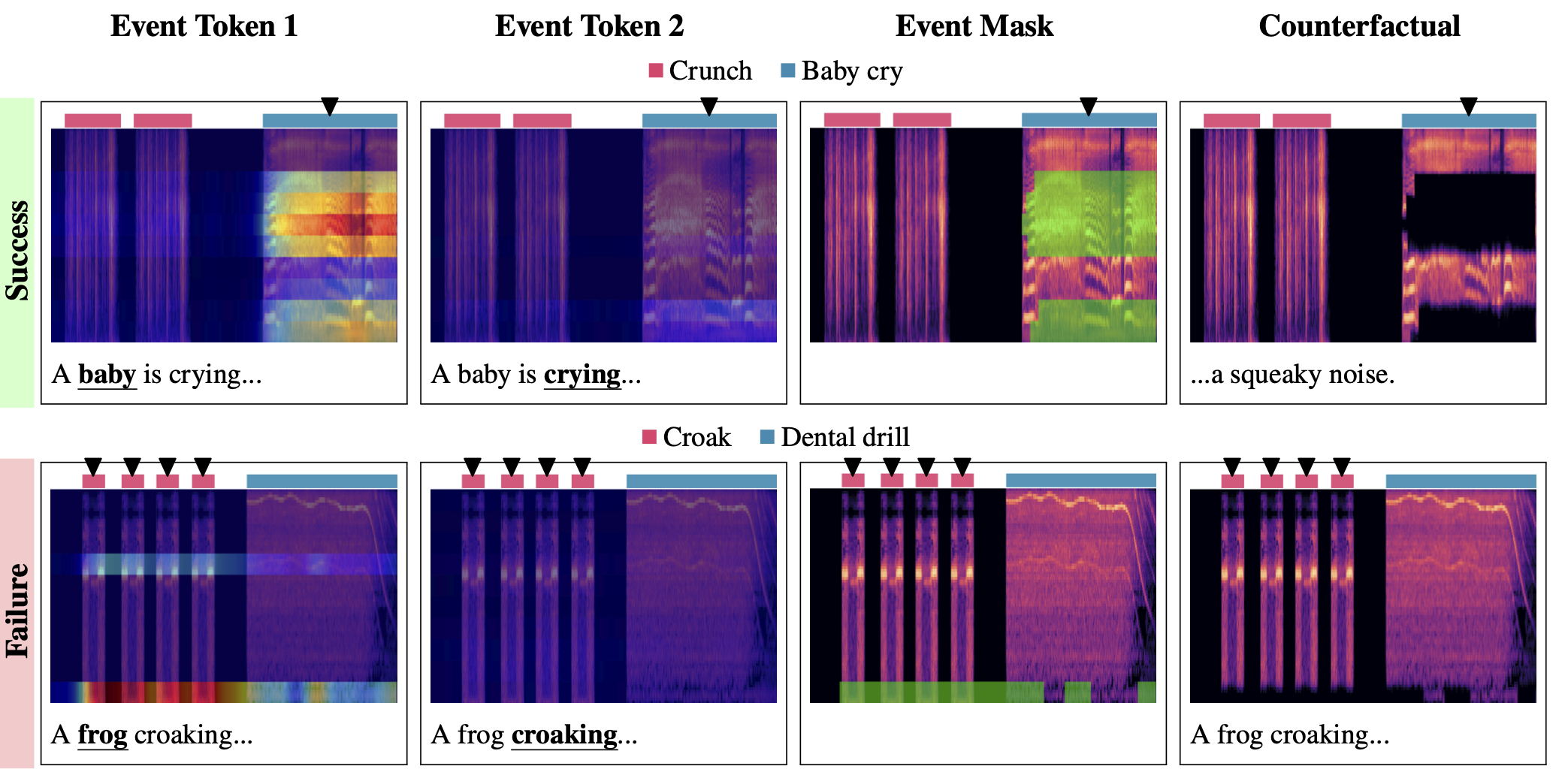}
    \caption{Representative counterfactual deletions. The first two columns show
the token maps associated with the target event, the third their aggregated
event mask, and the fourth the counterfactual spectrogram after deletion. In
the successful example, the target event disappears from the regenerated
caption; in the failure case, it remains. Green marks the selected deletion
region, while black indicates the removed time-frequency coefficients.}
    \label{fig:counterfactual}
\end{figure*}

\subsection{Counterfactual Faithfulness}
\label{sec:counterfactual-results}

We evaluate counterfactual deletion on 100 samples per benchmark, covering
$1{,}025$ generated acoustic events in total. For each event, we remove the
time--frequency region identified by \textsc{STAG} and measure how the
intervention affects the model output. The mean target-event log-probability
change is $\Delta\log p=-0.925$ (95\% bootstrap CI $[-0.993,-0.858]$;
$10{,}000$ resamples). Target confidence is reduced by at least one half in
$44.6\%$ of cases, and the event no longer appears in the regenerated caption
in $59.5\%$ of cases. Per-benchmark results are reported in the supplementary
material. Figure~\ref{fig:counterfactual} shows representative successful and unsuccessful
interventions. On average, the deletion masks $26.0\%$ of the time--frequency
plane and removes $37.3\%$ of the signal energy. Under the same intervention,
the mean log-probability change for non-target events is $-0.562$, compared with
$-0.925$ for the selected target. The larger effect on the target provides
behavioral evidence that the regions identified by \textsc{STAG} contribute
preferentially to the corresponding generated event, although overlapping
sources may still produce collateral effects.

\section{Conclusion}
\label{sec:conclusion}

We introduced \textsc{STAG}, a post-hoc framework for token-level
spectro-temporal grounding of captions generated by audio MLLMs. By combining
white-box temporal attribution with black-box Frequency-Band Occlusion,
\textsc{STAG} produces token-specific time-frequency relevance maps without
retraining or modifying the underlying model. These maps can be aggregated into
event-level explanations through a shared attribution and evaluation pipeline
applicable across heterogeneous audio-language architectures. Across four grounding benchmarks, combining temporal attribution with
frequency-band evidence yields more accurate grounding than ten post-hoc
explanation methods. The configuration analysis supports the complementarity
of spectral and temporal evidence, and shows limited sensitivity to the final
projection rule. Cross-model and qualitative analyses further reveal that
similar captions and temporal agreement can arise from different spectral evidence and source emphasis, while semantically plausible
label variations may be penalized by matching-based metrics. Counterfactual deletion provides behavioral support for the explanations:
removing the identified evidence reduces confidence in the corresponding target
events more strongly than in non-target events and frequently removes them from
regenerated captions. The current evaluation remains limited by the absence of
frequency-level annotations, ambiguity in benchmark labels, and the separable
fusion assumption used to construct each token map. Future work should therefore
develop spectro-temporally annotated benchmarks and investigate non-separable
attribution mechanisms capable of capturing localized time-frequency
interactions.

\bibliography{aaai2027}

\end{document}